# Thermal Management in Large Bi2212 Mesas used for Terahertz Sources

C. Kurter, K. E. Gray, J. F. Zasadzinski, L. Ozyuzer, A. E. Koshelev, Q. Li, T. Yamamoto, K. Kadowaki, W.- K. Kwok, M. Tachiki, and U. Welp

*Abstract*—We present a thermal analysis of a patterned mesa on a $Bi_2Sr_2CaCu_2O_8$ (Bi2212) single crystal that is based on tunneling characteristics of the c-axis stack of ~800 intrinsic Josephson junctions in the mesa. Despite the large mesa volume (e.g., 40x300x1.2 μm³) and power dissipation that result in self-heating and backbending of the current-voltage curve (I-V), there are accessible bias conditions for which significant polarized THz-wave emission can be observed. We estimate the mesa temperature by equating the quasiparticle resistance, $R_{qp}(T)$, to the ratio V/I over the entire I-V including the backbending region. These temperatures are used to predict the unpolarized black-body radiation reaching our bolometer and there is substantial agreement over the entire I-V. As such, backbending results from the particular $R_{qp}(T)$ for Bi2212, as first discussed by Fenton, rather than a significant suppression of the energy gap. This model also correctly predicts the observed disappearance of backbending above ~60 K.

## I. INTRODUCTION

A powerful and coherent electromagnetic radiation source at THz frequencies has been recently observed in relatively large mesa structures patterned on a $Bi_2Sr_2CaCu_2O_8$ (Bi2212) crystal [1]. Such emission from the natural stack of intrinsic Josephson junctions (IJJs) within the Bi2212 crystal structure resulted from a novel approach. An internal cavity resonance is used to achieve coherent Josephson oscillations [1] rather than the magnetic field used to excite the Fiske resonance

Manuscript received 19 August 2008. This work was supported by the US-Department of Energy, Basic Energy Sciences, under Contract No. DE-AC02-06CH11357, by JST (Japan Science and Technology Agency) CREST project, by the JSPS (Japan Society for the Promotion of Science) CTC program, by the Grant-in Aid for Scientific Research (A) under the Ministry of Education, Culture, Sports, Science and Technology (MEXT) of Japan and TUBITAK (Scientific and Technical Research Council of Turkey) project number 106T053.

C. Kurter is with the Materials Science Division, Argonne National Laboratory, Argonne, IL 60439 USA (phone: 630-806-3914; fax: 630-252-9595; e-mail: kurter@anl.gov).

L. Ozyuzer is with the Department of Physics, Izmir Institute of Technology, Izmir, 35430 Turkey.

J. F. Zasadzinski is with the Physics Division, BCPS Department, Illinois Institute of Technology, Chicago, IL 60616, USA.

K. E. Gray, A. E. Koshelev, Q. Li, W.- K. Kwok, and U. Welp are with Materials Science Division, Argonne National Laboratory, Argonne, IL 60439 USA.

T. Yamamoto and K. Kadowaki are with the Institute of Materials Science, University of Tsukuba, 1-1-1 Tennodai, Tsukuba-shi, Ibaraki-ken 305-8577, Japan.

M. Tachiki is with the Graduate School of Frontier Sciences, The University of Tokyo, 5-1-5 Kashiwanoha, Kashiwa 277-8568, Japan.

mode [2]. The large mesa size (e.g., 40x300x1.2 μm³) needed to obtain an internal cavity resonance at THz frequencies opposes the recent trend toward smaller mesas to reduce heating effects. Understanding the achievement of thermal management in such large mesas and its analysis is the theme of this paper.

The effect of self-heating on IJJ spectra is most evident in the backbending of the I-V (see Fig. 1). This shape is reminiscent of the slight reduction of the energy gap, $\Delta$, sometimes seen in low-resistance tunnel junctions of low-$T_c$ thin films at high currents. In these latter cases, the large current density results in self-heating and/or non-equilibrium quasiparticle injection effects [3,4]. The very large current density and close proximity of neighboring junctions in the crystal structure of Bi2212 could be expected to cause similar heating effects. Faced with this issue, researchers have found solutions to heating: use of short current pulses [5]; reduction of the current density by increasing the c-axis resistance through intercalation of inert molecules, e.g., $HgBr_2$, $HgI_2$, $I_2$ within the Bi-O bilayer of Bi2212 [6]; and reduction of the mesa volume which has been discussed in detail both experimentally [7] and theoretically [8].

The small voltage scale of the backbending, compared to the equilibrium energy gap (found, e.g., in high-resistance point-contact [9, 10] and STM [10, 11] tunneling, photoemission [12], etc.), presents a serious objection to the above picture of a simple reduction of the gap by heating. The reduction of $2\Delta$ in Fig. 1 to $\sim$2 meV=1.6 V/800 IJJs, from the equilibrium value of $\sim$100 meV for underdoped Bi2212 with $T_c\sim$100 K [9], would require a significant temperature rise that must be almost the same for heating powers that span a factor of about three in the negative resistance region. As this seems unrealistic, we pursue an alternative understanding, proposed by Fenton [13] and connected by Zavaritsky [14] to the temperature-dependent mesa resistance of Fig. 2, which we suggest is the quasiparticle resistance, $R_{qp}(T)$, for d-wave superconductors like Bi2212. In our picture, the voltages throughout the backbending I-V are always sub-gap, and we show below that backbending is a result of the particular shape of $R_{qp}(T)$ in Bi2212. Following this idea, we compare V/I with the $R_{qp}(T)$, determined separately, to find the mesa temperatures along the I-V: these temperatures follow Newton's law of heating, i.e., we find an approximately linear dependence on power dissipation, IV, that extrapolates back to the bath temperature, $T_0$, at zero current. Using only $R_{qp}(T)$ and Newton's law of heating our analysis correctly predicts: (1) the resistance of the mesa I-V's at the turning points of the backbending curve, i.e., dV/dI=0 in Fig. 1; and (2) the observed disappearance of backbending for $T_0$ above $\sim$60 K.

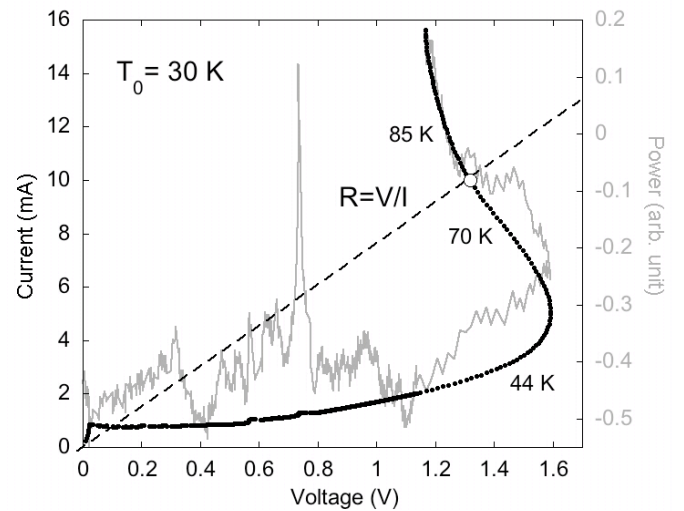

Fig. 1. Current-voltage (I-V) curve (dark line) for a mesa showing the commonly seen backbending for currents above ~5 mA. The listed temperatures are calculated in the text. The grey curve is the radiation simultaneously recorded by a remote bolometer with the peak at about 0.75 V being coherent radiation from synchronized intrinsic Josephson junctions. The dashed line illustrates the mesa resistance along the I-V.

## II. EXPERIMENT

Single crystals of Bi2212 were grown by the floating zone technique [15]. To obtain appropriately oxygen-deficient single crystals required a special annealing treatment: as-grown crystals were baked at 375 $^0$C in Ar flow (92 sccm) for 100 hours in a gettered furnace. The resulting underdoped crystals, which have a critical temperature of 78-82 K, are mounted on sapphire substrates using conductive epoxy. The typical contact resistance at the top of the mesa, of $\sim$10-30 ohms (compare to Fig. 2), is achieved by cleaving the Bi2212 crystal and then depositing Au to an eventual thickness of 100 nm. After that, mesas were fabricated using photolithography and Ar-ion milling. As there is only one contact on top of the mesa, we have a three-terminal measurement of the interlayer tunneling I-V and out-of-plane transport, $R_{qp}(T)$.

Mesas are mounted on a copper cold finger in a gas flow cryostat with a THz-transparent window directed at an IR bolometer. Copper tubes are used to channel the radiation and a mechanical chopper provides modulation for lock-in detection of the bolometer output to reduce background.

The analysis proceeds from separate measurements of I-V (Fig. 1) and $R_{qp}(T)$ (Fig. 2) from which the mesa temperature is determined. In addition, the far-field bolometer response is recorded simultaneously with the I-V (Fig. 1).

## III. ANALYSIS

The various steps in the analysis and its experimental validation are presented in the following sections.

### A. Quasiparticle Resistance, $R_{qp}(T)$, for $T < T_c$

The temperature-dependent quasiparticle resistance, $R_{qp}(T)$, is given by the c-axis resistance of a mesa if the Josephson supercurrent, $I_{cj}$, is suppressed. The zero-field data of Fig. 2 then define $R_{qp}(T)$ for $T > T_c$. To extend this below $T_c$, a field of B=10 T is used to suppress $I_{cj}$ down to 40 K by creating 2D pancake vortices throughout the ab-planes [16]. The thermal motion of these weakly pinned vortices destroys the ab-plane phase coherence of superconductivity. That prevents zero ab-plane resistance and severely shrinks the effective coherence area, A, for Josephson coupling between neighboring bilayers, and thus the Josephson energy, $E_J = hJ_cA/2\pi e$, can become much less than $k_BT$. Here $J_c$ is the intrinsic Josephson critical current *density* and A has been shown to be proportional to 1/B [17], so that large fields suppress $I_{cj}$ to even lower T. Although not critical to the following analysis, we extrapolate the data of Fig. 2 to lower temperatures in a manner consistent with the direct measurements of Yurgens, et al [18].

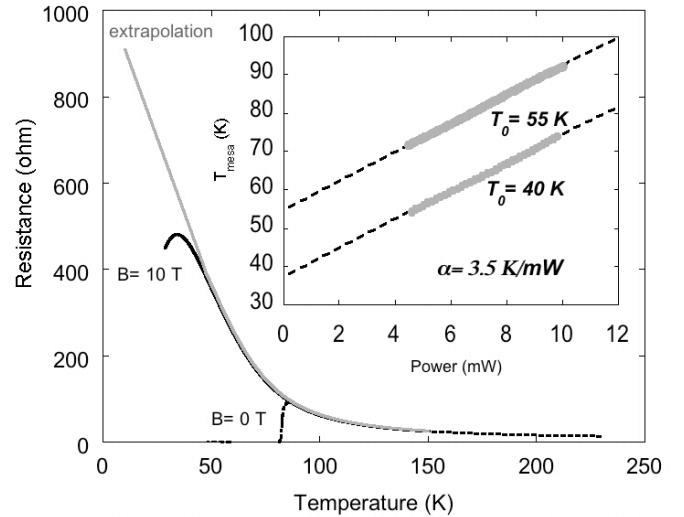

Fig. 2. Temperature dependence of the quasiparticle resistance, $R_{qp}(T)$, in zero field (T>$T_c$) and in a 10 T field down to $\sim$ 40 K and the extrapolation to lower T. The inset illustrates the mesa temperatures (grey line) determined from $R_{qp}(T)$ and V/I data of Fig. 1 as a function of power, P=IV. Dashed lines fit Newton's law of heating and extrapolate close to the bath value, $T_0$.

### B. Method to Determine the Mesa Temperature along I-V

Our method to determine the local temperatures, T, along an I-V curve, as in [13, 14], finds the value of T for which $R_{qp}(T)$ from Fig. 2 equals V/I from Fig. 1. Such temperatures are shown in Fig. 1, and we now present features of the results that give us confidence in the method's validity.

### C. Validating the Method

The mesa temperatures as a function of power dissipation, IV, are shown in the inset of Fig. 2 for bath temperatures, $T_0$, of 40 and 55 K. They exhibit an approximately linear dependence of T on power and, importantly, their extrapolation to zero power is quite close to the starting bath temperature, $T_0$. This physically reasonable behavior obeys Newton's Law of heating for a thermal resistance, $\alpha$,

$$T = T_0 + \alpha P = T_0 + \alpha I^2 R_{qp}(T). \qquad (1)$$

We find $\alpha\sim$3.5 K/mW, fairly independent of temperature, based on data sets at other $T_0$ over the range from 10 to 100 K.

We now show that, by using only $R_{qp}(T)$ and Newton's law of heating, our analysis correctly predicts: (1) the resistance of the mesa IV's at the turning points of the backbending curves, i.e., dV/dI=0 in Fig. 1 or 3; and (2) the observed disappearance of backbending for $T_0$ above $\sim$60 K. Finally, in Sec. D, we show that these mesa temperatures provide substantial agreement with the unpolarized blackbody radiation reaching our bolometer over the entire I-V.

The turning points in the backbending I-V can be calculated by solving dV/dI=0 subject to the constraint of Eq. 1, giving

$$dR_{qp}(T)/dT = R_{qp}(T)/(T_0-T),\qquad(2)$$

$$R_{qp}(T) = c/(T-T_0).\qquad(3)$$

For backbending, the hyperbola, $c/(T-T_0)$, will be tangent to $R_{qp}(T)$ for two values of c, since the I-V curve must have two turning points to eventually have a positive slope at very large current. The first turning points are readily visible in the I-V data of Fig. 3, which is shown for a wide range of $T_0$. The determination of the tangent condition is illustrated in Fig. 5 for $T_0$=45 K. There are two values of c giving tangents at ~73 K and ~125 K and at the first turning point, $R_{qp}$~172 Ω. This is to be compared with the turning-point resistance at 45 K (circle in Fig. 3) that is ~184 Ω. In Fig. 4, the procedure is inverted to calculate I-V's using no adjustable parameters, since Fig. 2 gives α and $R_{qp}(T)$. The larger deviations at lower $T_0$ are possibly due to the inadequate extrapolation of $R_{qp}(T)$.

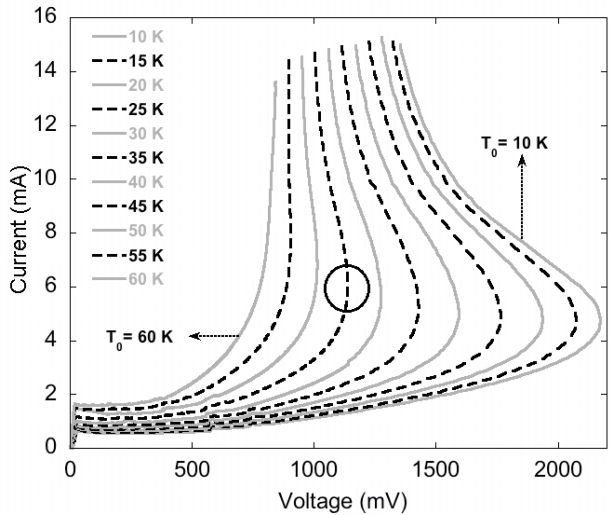

Fig. 3. Temperature dependence of mesa I-V showing back bending (turning points with dV/dI=0) at low temperatures that disappear for $T_0 \geq 60$ K. The turning point at 45 K is indicated for a comparison described in the text.

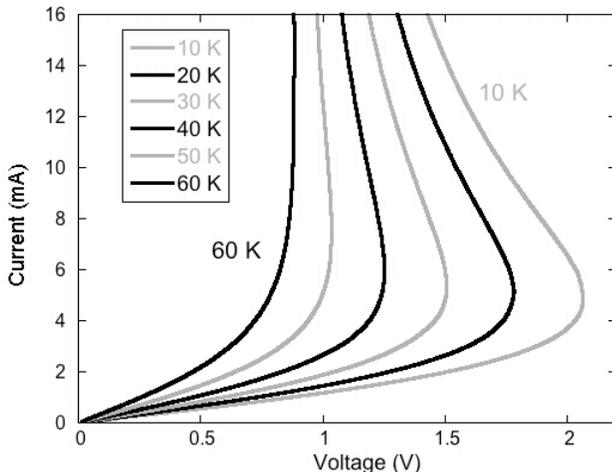

Fig. 4. Using no adjustable parameters, we show the calculated I-V's corresponding to the data of Fig. 3. Heat transfer of 3.5 K/mW comes from inset of Fig. 2 and $R_{qp}(T)$ from main body of Fig. 2.

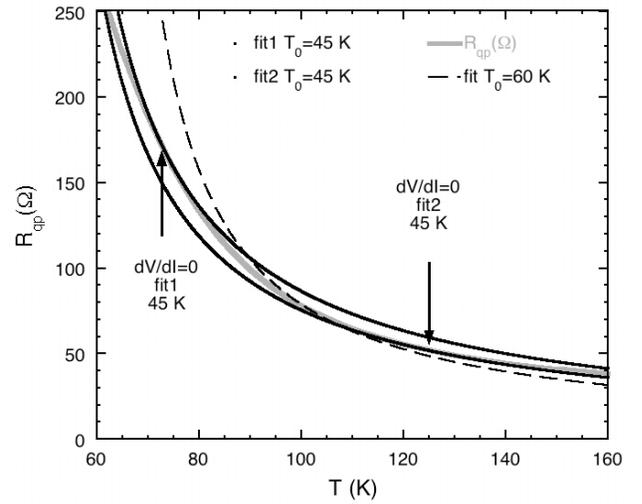

Fig. 5. Temperature dependence of quasiparticle resistance data (grey line) and tangent fits to the hyperbola of Eq. 4. For $T_0$=45 K there are tangents at ~73 K and ~125 K for different values of the constant c in Eq. 4 implying the I-V would show back bending. For $T_0$=60 K, no tangent fit to Eq. 4 can be found for any value of c, i.e., every hyperbola crosses the data (dashed line).

The dashed line in Fig. 5 shows that for $T_0$=60 K, it is not possible to find a parabola that is tangent to $R_{qp}(T)$. This implies no turning point and no backbending as is amply demonstrated in Fig. 3 for $T_0$=60 K. We find that the loss of backbending at ~60 K is commonly seen in our Bi2212 mesas as well as ones in which Bi2212 has been intercalated with HgBr₂ to greatly reduce the dissipation by greatly increasing the c-axis resistance and $R_{qp}(T)$. Thus backbending seems to depend on shape of $R_{qp}(T)$ but less on its magnitude.

### D. Compare Unpolarized Bolometer Response with Black-Body Calculation from Mesa Temperature

The bolometer response is shown in Fig. 6 for a different mesa. To verify that the high-voltage unpolarized radiation results from the mesa temperatures found, as in Sec. B, we calculate the expected blackbody radiation from the Bi2212 surfaces facing the bolometer. The emitted power collected within the solid angle for the bolometer, Ω~π/2, is

$$P_B = eS(\Omega/2\pi)\sigma T^4,\qquad(4)$$

where e is the emissivity (assumed to be 1), S is the emitting area and σ is the Stephan-Boltzmann constant. Our bolometer has a low-pass filter to suppress background radiation for wavelengths less than ~120 microns. The calculation of the effect of this filter on the black-body spectrum at temperature T results in replacing $T^4$ in Eq. 4 by a response function, F(T) that is 0.8 $T^4$ at low temperatures and approximates $6.4 \times 10^4$ $K^3$ (T-27.5 K) for the range from 40 to 150 K. Using the calculated mesa temperatures, we arrive at the solid line fit to the high-voltage bolometer response shown in Fig. 6. The shape agrees quite well, and the magnitude of the bolometer response agrees within ~30% if blackbody radiation from the *entire crystal* is included. The area of the Bi2212 crystal exposed to the bolometer is ~700 times greater than the mesa alone and because of the linear dependence of F(T) on T, even modest temperature increases in the whole crystal can swamp the contribution of the hotter mesa. The 3D heat-flow equation for the Bi2212 crystal/mesa combination is solved using the appropriate geometry and literature values for its anisotropic thermal conductivity [19]. This solution indicates an *average* temperature for the exposed crystal wall that is ~15% of the mesa itself. Including this contribution, the agreement of the magnitude within 30% further validates the use of $R_{qp}(T)$ to determine the mesa temperature.

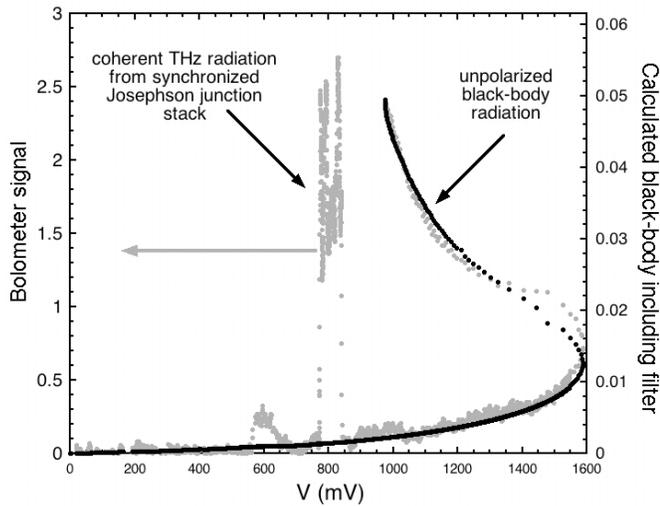

Fig. 6. The bolometer response (grey points) at 26 K for a 60-μm wide mesa shows peaks due to coherent radiation at ~800 mV and an unpolarized high voltage thermal source. Use of T, determined from our method (Sec. B), adequately predicts the shape of the scaled blackbody data at these high voltages (solid line) and it is within 30% in magnitude.

## IV. CONCLUSION

We demonstrate an understanding of the thermal management that has been achieved in relatively large mesas on Bi2212 crystals that emit considerable power in the THz frequency range [1]. In addition, we show that the backbending universally seen in the I-V of Bi2212 mesas has its origins in the particular temperature dependence of $R_{qp}(T)$, rather than a significant suppression of the energy gap.

## ACKNOWLEDGMENT

We thank A. Imre at the Center for Nanoscale Materials, Argonne National Laboratory, for help with the electron microscopy and FIB work.